\documentclass{ws-p8-50x6-00}
\usepackage[dvips]{epsfig}
\begin{document}

\title{MULTIPLE PRODUCTION, TRANSPORT IN ATMOSPHERE \\
  AND DETECTION OF HIGH ENERGY COSMIC RAYS}

\author{Jean-No{\"e}l Capdevielle,
Fabrice Cohen 
and Corentin Le Gall
}

\address{PCC,Coll{\`e}ge de France, 11, Place Marcelin Berthelot,75231
  Paris C{\'e}dex 05 France\\
  E-mail: capdev@cdf.in2p3.fr}

\author{
Izabela Kurp,
Barbara~Szabelska and Jacek Szabelski}

\address{
The Andrzej So{\l}tan Institute for Nuclear
  Studies,
  90-950 L{\'o}d\'z 1, Box 447, Poland\\
  E-mail: js@zpk.u.lodz.pl}


\maketitle

\abstracts{We describe the general aspects of Monte Carlo Collision
  Generators suitable for cosmic ray nucleon-Air and nuclei-Air
  interactions, including accelerator and collider data. The problem
  of the extrapolation at 3 energy decades above the LHC of the main
  features of high energy collisions is discussed and under
  theoretical and phenomenological assumptions, the properties of the
  longitudinal and lateral development of giant extensive air showers
  simulated with the CORSIKA program are presented. The determination
  of the primary energy near $10^{20}$~eV is examined for different
  observables, total size, densities of charged particles interpolated
  at 600~m from shower core.\\
  The extensive air shower data collected
  around LHC energy is in better agreement with models of large
  multiplicities. Beyond this energy, the extrapolation carried
  assuming the diquark breaking mechanism can change the classic
  conversion to primary energy and such circumstance can have
  consequences on the validity of the GZK cut off.\\
  In those conditions, we have simulated large and giant air showers
  taking into account, in addition, new processes, such as diquark
  breaking, and topological problems involving adequate structure
  functions for lateral distributions, up to energies exceeding
  $10^{20}$~eV for P.AUGER and EUSO experiments.}

\section{Monte Carlo Collision generators for Cosmic Rays 
  and extrapolation at ultra high energy} 
\begin{figure}[t]
  \epsfxsize=7cm 
  \epsfbox[0 0 567 600]{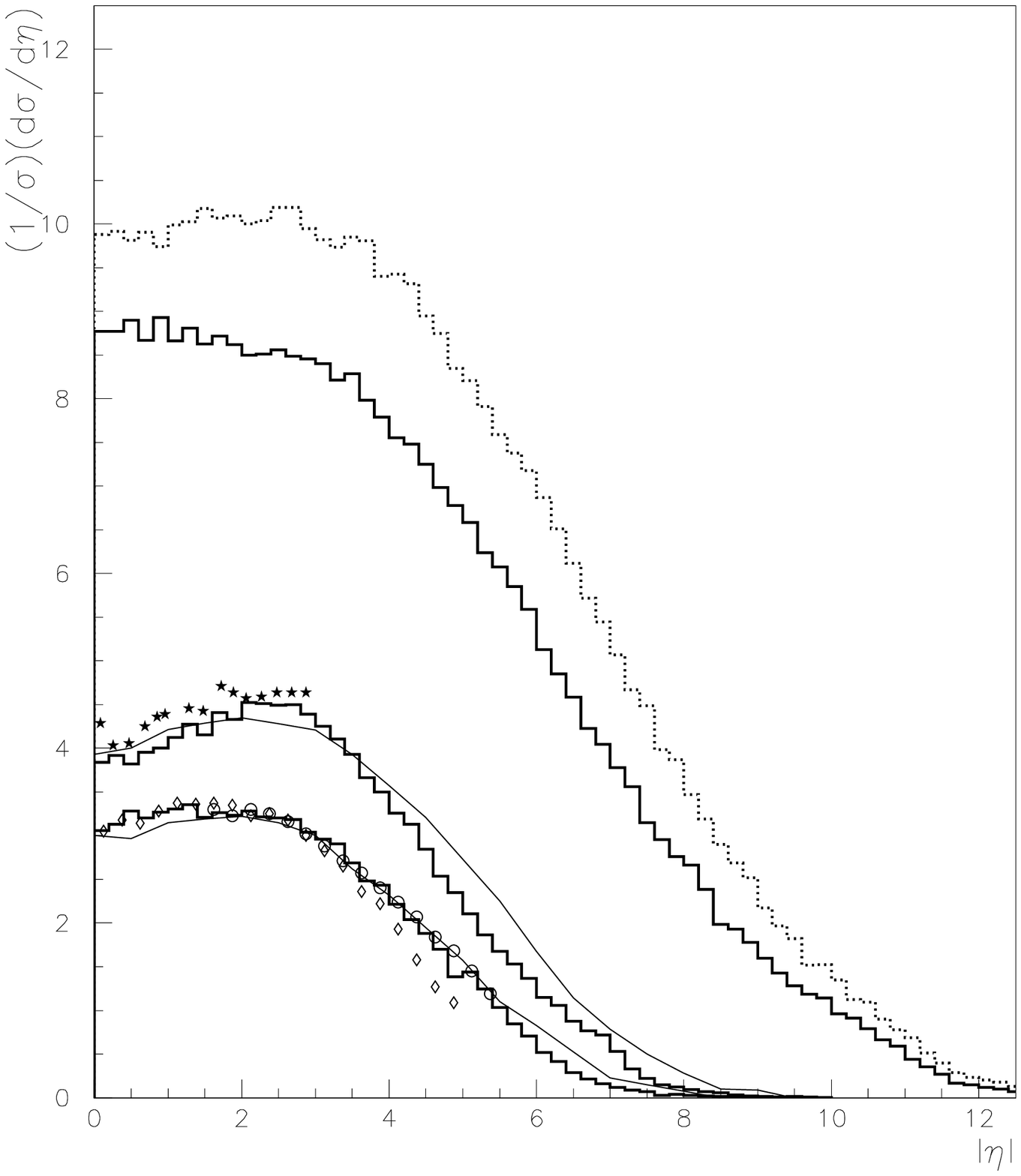}
  \epsfxsize=4cm 
  \epsfbox[99 530 400 736]{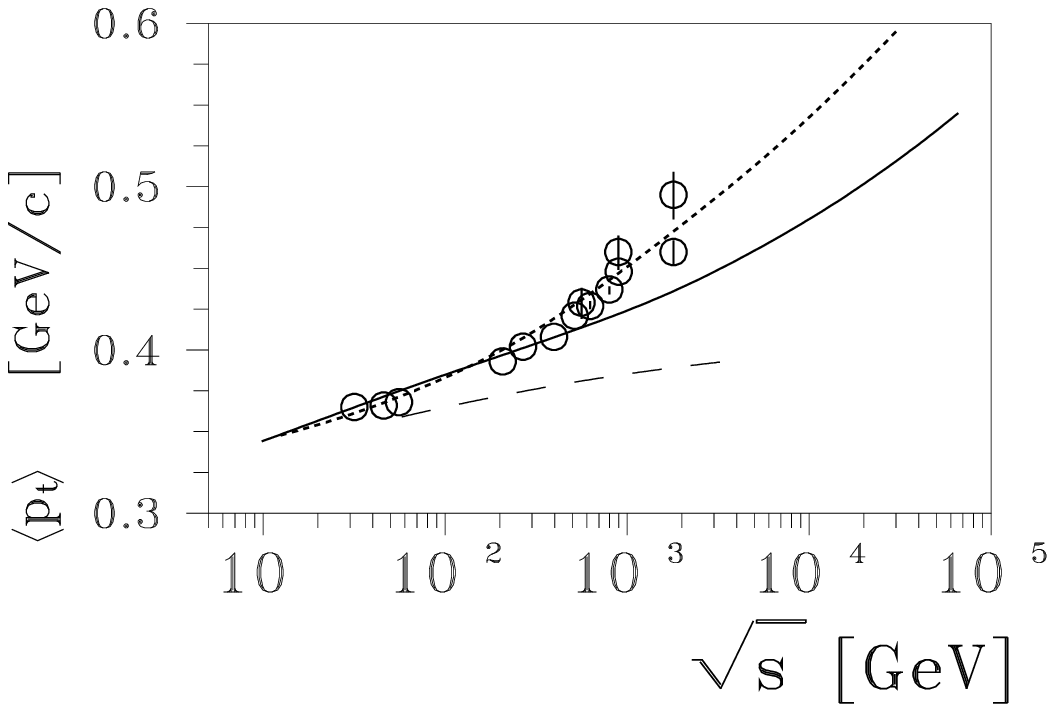}
  \caption[ps and pt]
  {Left:
     pseudo--rapidity distribution with HDPM2. \label{fig:PSdis}\\
  Right:
  average $p_{t}$ vs. energy  dependence with HDPM2.\label{fig:PSdis4}} 
\end{figure}

In order to simulate in a
reasonable time the 4-dimensional development of Extensive Air Showers
at very high energy, it is important to elaborate collision generators
reproducing rapidly the detailed features of multiple production
observed in accelerators and colliders. Among the variety of models
implemented in CORSIKA, there are microscopic and
phenomenological models. The first ones include all the steps
of the parton momenta generation from the parton distribution
functions related to valence quarks and diquarks, sea quarks and
gluons, convoluted with the fragmentation of the respective strings
into secondary hadrons. The phenomenological models provide fast and
direct hadron sampling, but take into account some global
characteristics of the Gribov-Regge theories with parameters adjusted
to the experimental data. One common feature is the separation between
the non single diffractive component and the diffractive component
(single and double) with the
inclusion of soft and hard mechanisms.\\
As an example (fig.\ref{fig:PSdis}), the reproduction of the
inclusive data with the model HDPM2 (2nd version of the hybrid dual
parton model) with parameters tuned to fit the experimental data of
FermiLab
 at $\sqrt s = 630 GeV$ gives a total
average inelasticity for p-p collisions of $0.7$ instead
of $0.5$ when adjusted to the previous measurements of UA5.
The forward trajectory on Fig.~\ref{fig:PSdis} is not constrained
above 5.5 units of pseudo rapidity and several models taken in option
for CORSIKA can have different trajectories in the forward region with
a good agreement to the observed data, but rather different
inelasticities.  A cosmic ray cascade is initiated by the unique
interaction of the primary particle and the fluctuations of the
elementary act have to be reproduced carefully taking into account the
semi-inclusive data 
 as well as the correlation between
charged and neutral secondaries.
The extrapolation at ultra high energy can be carried with the model
HDPM2 taking into account recent features of collider physics such as
$ p_{t}$ versus central rapidity density (UA1-MIMI exp.)  and recent
results of FermiLab~\cite{ref:JNC99}
 for pseudo-rapidity up to 5.5.
The pseudo-rapidity distributions obtained~\cite{ref:JNC00}
 with HDPM2 for 2000
collisions NSD are shown on Fig.~\ref{fig:PSdis}.
The distributions on Fig.~\ref{fig:PSdis} (histograms for HDPM2, full
line for QSJET model) give for HDPM2 different extrapolations at
$10^{10}$ GeV following the PDF
assumed ($B0$ and $B_{-}$).\\
The correlation between the central rapidity density and the average
transverse momentum $<p_{t}>$ plays also one role in the muon and
hadron radial distributions: the dependence of the transverse
momentum on
energy turns to a permanent increase on Fig.~\ref {fig:PSdis4}.

The p-p collision is transformed to p-A collision following 
Glauber's considerations and Nuclei-Air collisions are treated by the
abrasion evaporation procedure.

\begin{figure}[t]
  \epsfxsize=5.8cm 
  \epsfbox[0 0 550 540]{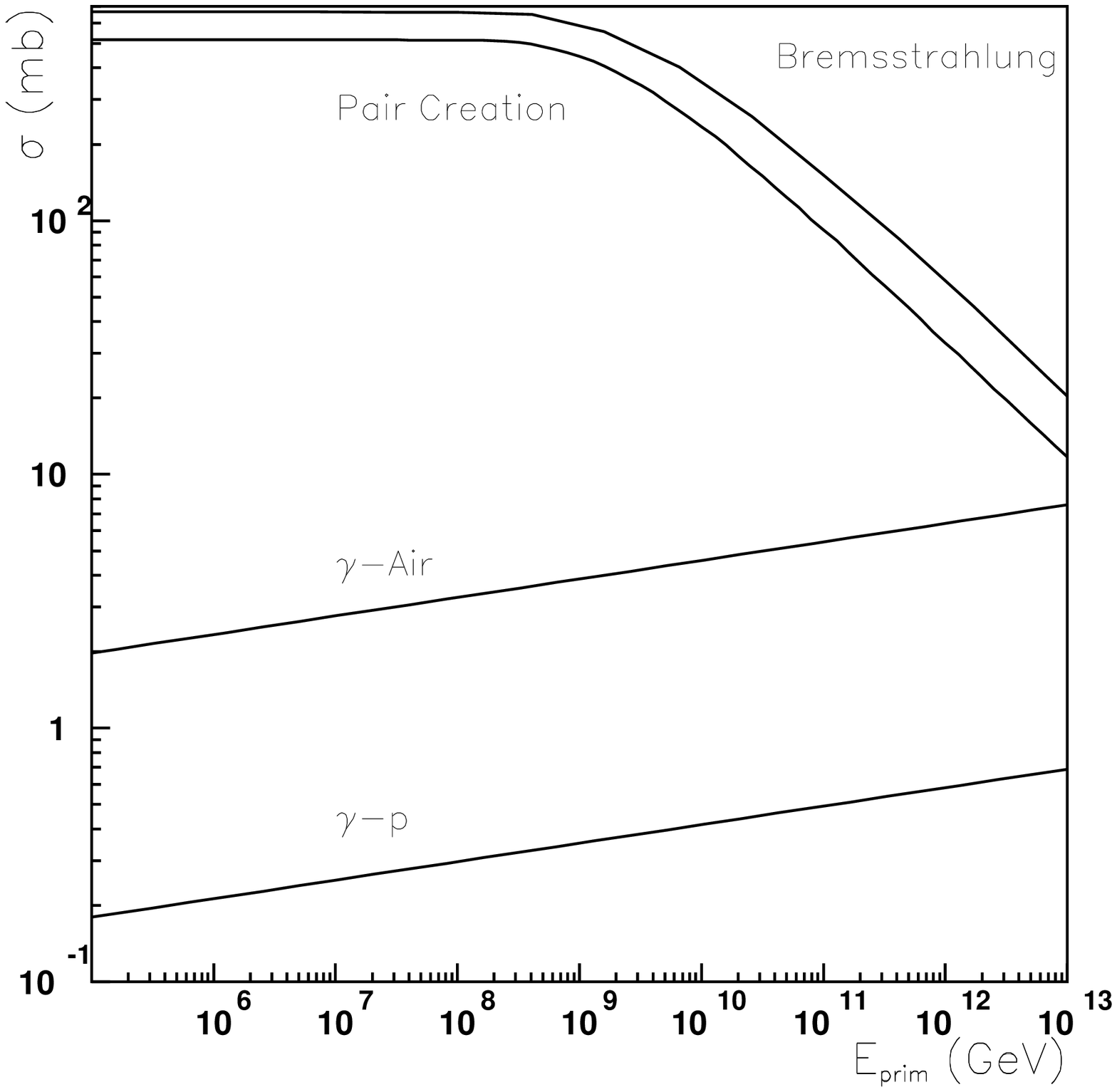}
  \epsfxsize=5.8cm 
  \epsfbox[0 0 530 500]{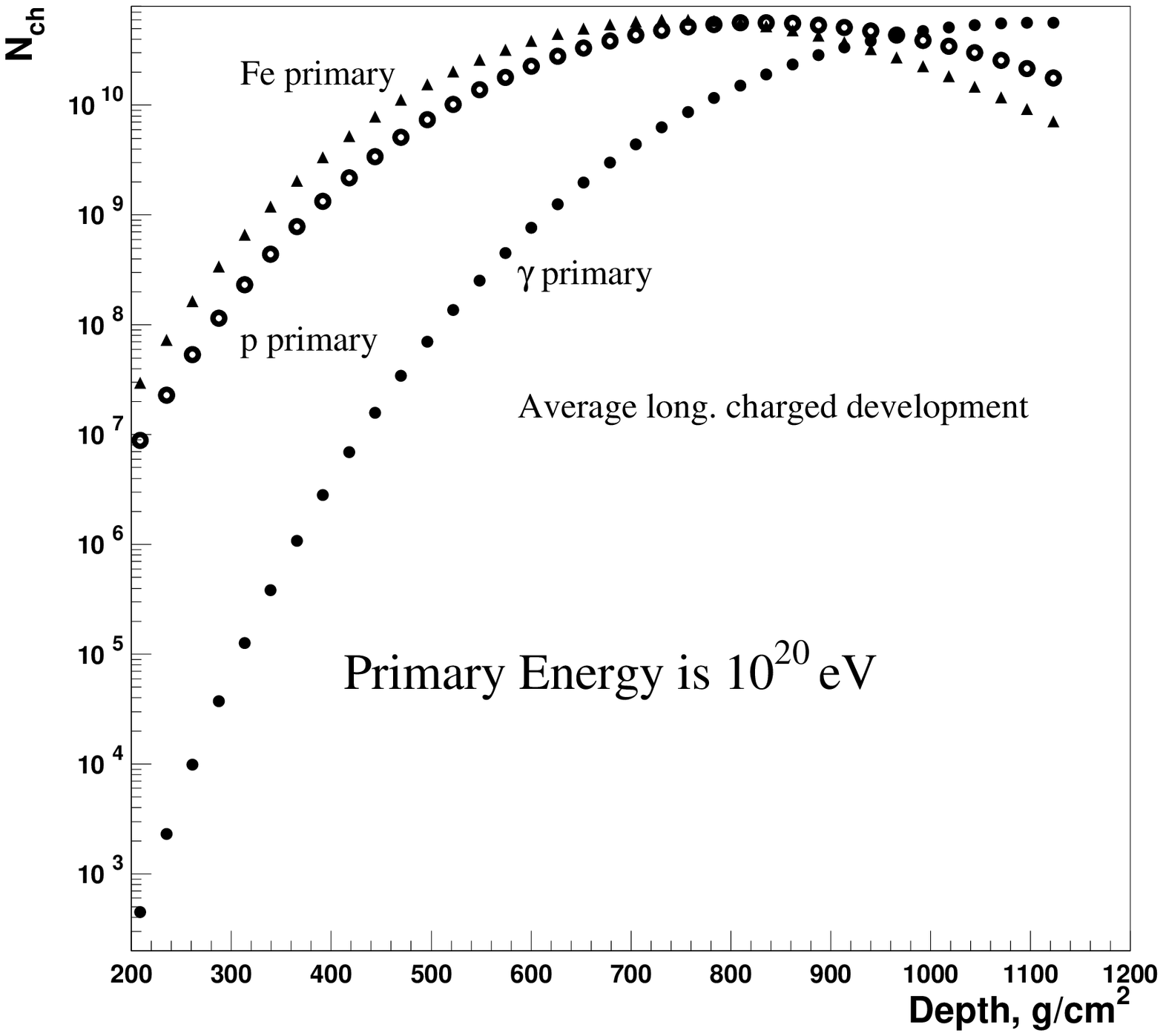} 
    \caption[photo-pr + long]
   {Left:
      Photo--production cross section in air compared with 
      bremsstrahlung and pair production cross sections reduced by the LPM
      effect. \label{fig:PSdis5}\\
   Right:
      longitudinal development of EAS for $\gamma$, p and Fe primaries.
      \label{fig:PSdis6}}
\end{figure}


\section{Simulation of giant EAS}

The electromagnetic cascade is treated by EGS4 involving cross
sections, taking into acount the LPM effect, for bremsstrahlung and
pair production. The nuclear photo production has been implemented in
the simulation and its contribution is compared versus the primary
photon energy to the previous processes on the
Fig.~\ref{fig:PSdis5}~\cite{ref:JNC00}.

The fluctuations of multiplicity are governed by the negative binomial
distribution and at $10^{20} eV$, the average charged multiplicity
is close to 300 for HDPM (1000 for QGSJET); this corresponds to
energy densities of $ 30~GeV/fm^{3}$ in average... and we have not considered
here the important modifications from a possible phase transition to
QGP.\\
 The longitudinal developments obtained for proton, photon and
iron induced extensive air showers are compared on
Fig.~\ref{fig:PSdis6}. We notice that the showers (here averaged on 10
individual cascades simulated with CORSIKA) have their maximum near
1000--1500~m altitude for proton and iron. The $\gamma$ shower initiated
at $50~g/cm^{2}$ exhibits a different behaviour due to the LPM
effect; a rule of thumb for hadronic showers is to divide by~2 the
total energy in $GeV$ to get $ N_{max}$. 
This maximum depth 
corresponds to a minimum of fluctuations, suggested by analytic 
cascade theory and confirmed by the Monte Carlo simulation, indicating
the well adapted localisation of AGASA and AUGER experiments. The 
depth of the maximum depends on the logarithm of the primary energy, 
when $N_{max}$ remains proportional to energy 
with a similar factor for a large variety of interaction models.\\
A more delicate problem is the estimation of $N_{max}$; it can be 
read directly from the cascade curve derived from fluorescence 
measurements, as performed by the Fly's Eye or Hires, but the approach 
with a small number of detectors hit at large distances (1 to 1.5 km)
 turns to a hopeless topological problem.

\section{Topological aspects of radial distribution}

The profile of the lateral distribution assumed and the method of core
localization are especially important. We propose instead of NKG and
other Euler Beta functions, the employment of the gaussian
hypergeometric formalism giving also normalization and better
skewness, under the form:
\begin{eqnarray}
  f(x)
  &= g(s)x^{s-a}(x+1)^{s-b}(1+dx)^{-c} \label{eq:JNC}
\end{eqnarray}
which has the advantage (for values of parameters
respecting the conditions of convergence $s-a+2>0$ and $c-2s+b-2>0$)
to be exactly normalized in terms of
Gaussian Hypergeometric function $F_{HG}$~$=$~$F(c,s-a+2,c+b-s;1-d)$ by:

\begin{equation}
  \ g(s) = \frac{ \Gamma(c+b-s)}{2\pi\Gamma(s-a+2)\Gamma(c-2s+b+a-2)
    F_{HG}}
  \label{eq:g(s)}
\end{equation}

The empirical distributions, such as AGASA function~\cite{ref:Nag92},
 as underlined by Vishwanath~\cite{ref:Vis}
 enters in the category of Hypergeometric Gaussian functions
~\cite{ref:CP}, under the general form of structure function
\begin{equation}
  \ f(x) =  C_{ e} \cdot x^{ - \alpha } \cdot
  (1+x)^{ - ( \eta - \alpha )} \cdot
  ( 1+ dx ) ^{ -\beta}\\
  \label{eq:AGASAHG}
\end{equation}
with the conditions $2-\alpha>0$ and $\beta+\eta-2>0$.
The value used in AGASA function
for the coefficient $C_{e}$ is just an approximation; the exact value
is
\[
C_{ e} = \frac{ \Gamma (\beta+ \eta - \alpha ) } { 2 \pi
  \cdot \Gamma ( 2 - \alpha ) \cdot \Gamma (\beta+ \eta - 2 )} \cdot
\frac{ 1}{ F_{HG}}\] where $F_{HG}=
F_{HG}(\beta,2-\alpha,\beta+\eta-\alpha;1-d)$. The Hypergeometric
Gaussian function can be easily calculated from the hypergeometric
series:
\begin{eqnarray*}
  \ F_{HG}(a,b,c;z) =
  \sum_{n=0}^{\infty}\frac{(a)_{n}(b)_{n}}{(c)_{n}n!}z^{n},
  \rule{0.2cm}{0cm} c\neq 0, -1, -2, ..., \\
  (a)_{n}=\Gamma(a+n)/\Gamma(a),
  \rule{0.2cm}{0cm} (a)_{0}=1
\end{eqnarray*}
This equation is equivalent to our version $(3)$ containing the age parameter s with the relations between respective coefficients:
\rule{0.4cm}{0.cm}
$ x=\frac{ r}{ r_{\rm M}}$,
\rule{0.4cm}{0.cm}
$ d =\frac{ r_{\rm M}}{ r_{0}}$,
\rule{0.4cm}{0.cm}
$ s = 1.03$,
\rule{0.4cm}{0.cm}
$ \alpha = a - s$,
\rule{0.4cm}{0.cm}
$ \eta = b - s + \alpha $ (the value of s is taken from the
longitudinal development simulated).

\begin{table}[t]
  \caption{Best parameters to simulated $e^{+}e^{-} +$ muons 
    (all charged) lateral distribution fit using JNC01 formula.
    \label{tab:JNCHGa}}
  \begin{center}
    \footnotesize
    \begin{tabular}{|c|c|c|c|c|l|}
      \hline
      {} &
      \raisebox{0pt}[13pt][7pt]{p10} &
      \raisebox{0pt}[13pt][7pt]{p20} & 
      \raisebox{0pt}[13pt][7pt]{Fe10} & 
      \raisebox{0pt}[13pt][7pt]{Fe20}\\
      \hline
{$ log_{10}N_{e}$} &
{10.75} &
{10.72} & 
{10.70} & 
{10.65}\\
{$r_{\rm M}$} &
{21.26} &
{21.26} & 
{19.18} & 
{19.18}\\
{$r_{0}$} &
{8785.} &
{8785.} & 
{9536.} & 
{9536.}\\
{$a$} &
{1.91} &
{1.91} & 
{1.82} & 
{1.82}\\
{$s$} &
{1.03} &
{1.04} & 
{1.03} & 
{1.04}\\
{$b$} &
{3.32} &
{3.32} & 
{3.31} & 
{3.31}\\
{$\beta$} &
{10.0} &
{10.0} & 
{10.0} & 
{10.0}\\

      \hline
    \end{tabular}
  \end{center}
\end{table}

\begin{figure}[h]
 \noindent
  \epsfxsize=5.8cm 
  \epsfbox[80 120 510 770]{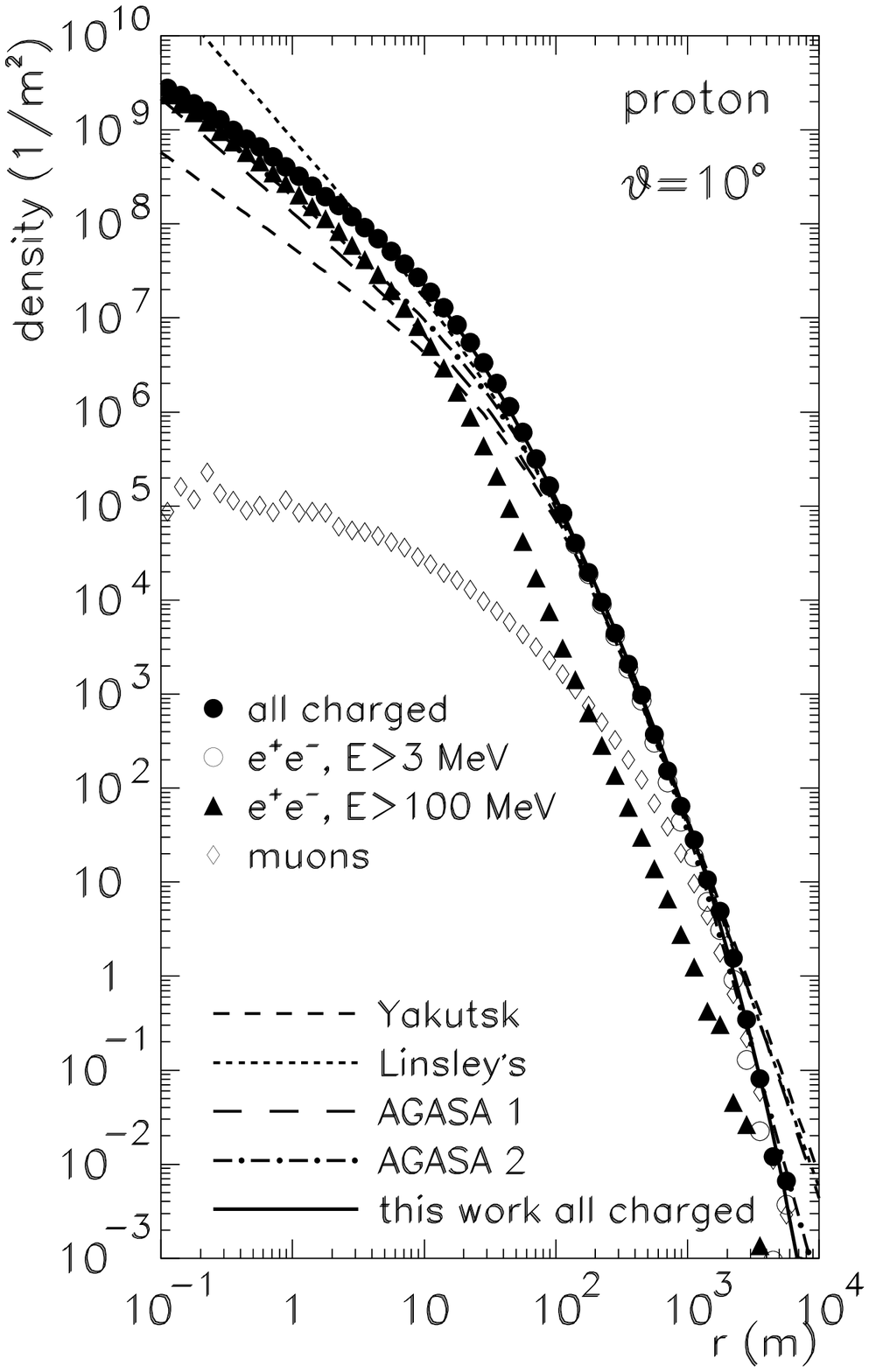}
  \epsfxsize=5.8cm 
    \epsfbox[87 300 550 800]{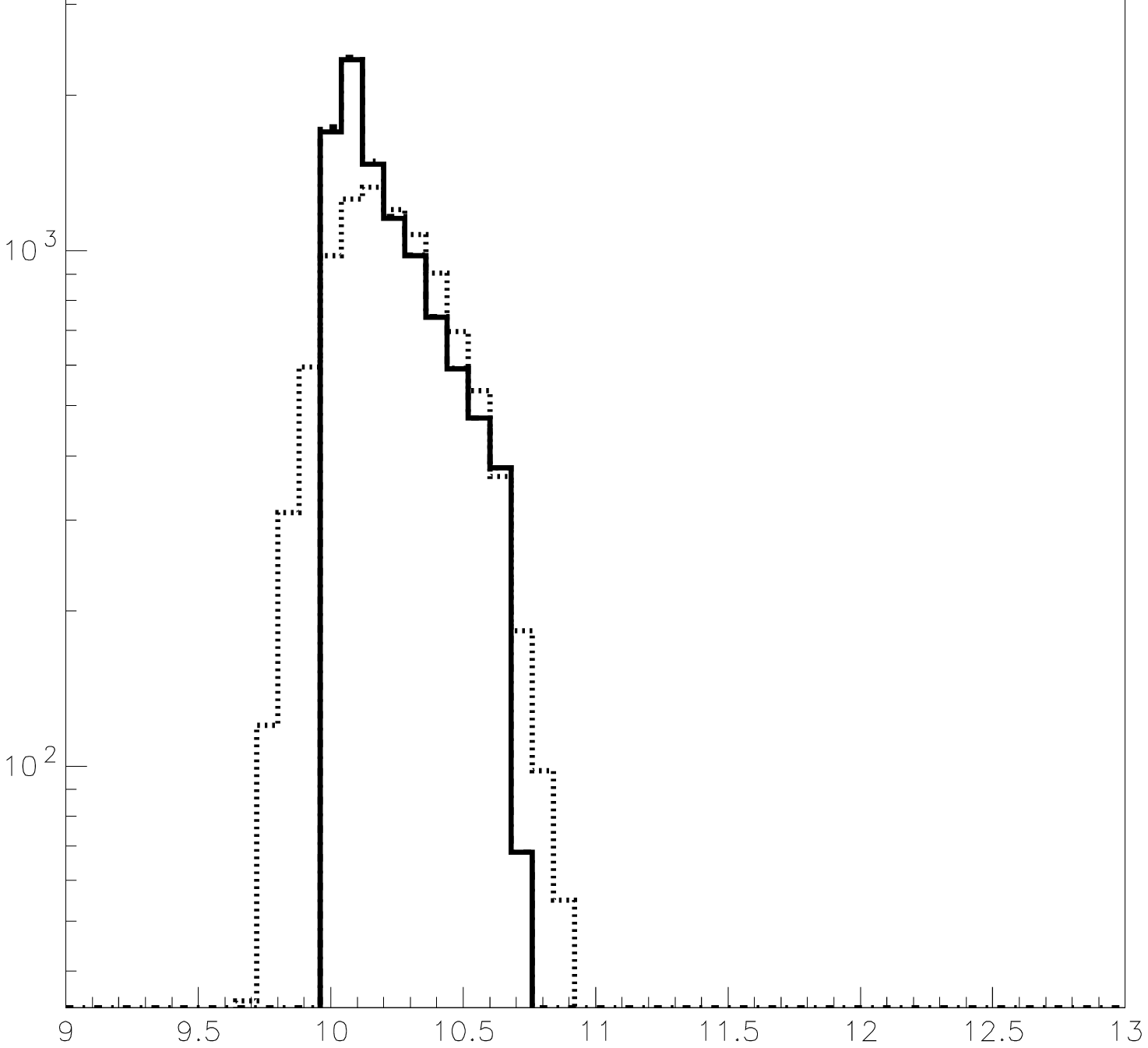} 
  \caption[two figures]
  {Left:
  fits to all charged particles lateral distribution from simulations
    (average from 10 EAS). Primary particle is a proton with
    energy 10$^{11}$~GeV.
    Lines are normalized to $\varrho$(600~m).\label{fig:JNCHGa}\\
  Right:
    extra component to GZK's cut off in case of diquark breaking.
      Full line: generated spectrum, dashed: reconstructed, versus
      $log_{10}(E_{0}/1~GeV)$ \label{fig:PSdis8} }
\end{figure}

\noindent
We have adjusted with MINUIT the parameters of our hypergeometric
function (Table~\ref{tab:JNCHGa}) to the average lateral distributions
(set JNC01 for charged, JNC02 for electrons) of groups of 10 showers
at $10^{20}$~eV simulated with CORSIKA (QGSJET model),
as shown on Fig.~\ref{fig:JNCHGa}.

In each case, the adjustment has been performed with 50 points from
the simulation distributed from 0.1~m up to 10~km from axis position
for charged particles
(muons and electrons). The advantages of JNC01 formula can be seen on Fig.~\ref{fig:JNCHGa}
and on Table~\ref{tab:comp}.

\begin{table}[t]
  \caption{Columns $a$ present total number of charged particles $N_{e}$ in
    10$^{10}$, columns $b$ the ratios $E_{0}/N_{e}$ in GeV
    ($E_{0}$=10$^{11}$~GeV) and columns $c$ the ratios
    $\varrho(600)/N_{e}$ in 10$^{-8}$ particles/m$^{2}$.  $m(600) =
    \frac{ d(log~\varrho)}{ d(log~r)}$ at 600 m.
    \label{tab:comp}}
  \center
  \footnotesize
  \begin{tabular}{||l||r|r|r||r|r|r||r|r|r||r|r|r||}
    \hline
    &
    \multicolumn{3}{c||}{
      proton 10$^{\circ}$
      }  &
    \multicolumn{3}{c||}{
      proton 20$^{\circ}$
      } &
    \multicolumn{3}{c||}{
      iron 10$^{\circ}$
      }&
    \multicolumn{3}{c||}{
      iron 20$^{\circ}$
      } \\
    \hline
    $\varrho(600)$ &
    \multicolumn{3}{c||}{
      290 m$^{-2}$
      }  &
    \multicolumn{3}{c||}{
      318 m$^{-2}$
      } &
    \multicolumn{3}{c||}{
      369 m$^{-2}$
      }&
    \multicolumn{3}{c||}{
      356 m$^{-2}$
      } \\
    \hline
    \parbox[c]{1.5cm}{
      \rule[0.cm]{0.cm}{0.1cm} $E_{0}/\varrho(600)$ \\
      (GeV~m$^{2}$) \rule[-0.4cm]{0.cm}{0.5cm}} &
    \multicolumn{3}{c||}{
      3.4 $\cdot$ 10$^{8}$
      }  &
    \multicolumn{3}{c||}{
      3.1 $\cdot$ 10$^{8}$
      } &
    \multicolumn{3}{c||}{
      2.7 $\cdot$ 10$^{8}$
      }&
    \multicolumn{3}{c||}{
      2.8 $\cdot$ 10$^{8}$
      } \\
    \hline
    $m(600)$ &
    \multicolumn{3}{c||}{
      --3.9
      }  &
    \multicolumn{3}{c||}{
      --3.6
      } &
    \multicolumn{3}{c||}{
      --3.6
      }&
    \multicolumn{3}{c||}{
      --4.0
      } \\
    
    \hline
    \hline
    fit &
    $1a$ & $1b$ & $1c$ &
    $2a$ & $2b$ & $2c$ &
    $3a$ & $3b$ & $3c$ &
    $4a$ & $4b$ & $4c$ \\
    \hline
    Yakutsk &
    1.8  & 5.6 & 1.6  &
    1.7  & 5.9 & 1.9  &
    2.3  & 4.3 & 1.6  &
    1.9  & 5.3 & 1.8  \\
    
    Linsley's &
    8.2 & 1.2 & 0.3 &
    8.0 & 1.3 & 0.3 &
    10.5 & 0.9 & 0.3 &
    8.9 & 1.1 &  0.4 \\
    
    AGASA\#1 &
    2.4 & 4.2 & 1.2 &
    2.6 & 3.8 & 1.2 &
    3.1 & 3.2 & 1.1 &
    2.9 & 3.4 & 1.2 \\
    
    AGASA\#2 &
    3.3 & 3.0 & 0.8 &
    3.6 & 2.8 & 0.8 &
    4.2 & 2.4 & 0.8 &
    4.0 & 2.5 & 0.8 \\
    
    this work &
    5.6 & 1.8 & 0.5 &
    5.6 & 1.9 & 0.6 &
    5.1 & 2.0 & 0.7 &
    4.5 & 2.2 & 0.7 \\
    
    \hline
  \end{tabular}
\end{table}

The major part of the particles is contained inside 200~m from the
axis and only the skewness of the hypergeometric function allows a
reliable relation between size and density at 600~m. This function has
been applied to localizations
of the showers contained in the catalogues of Volcano
Ranch and Yakutsk.  The core position has been obtained by
minimization with Minuit program between different formulas available
for lateral densities written versus the coordinates $X$, $Y$ as
\begin{equation}
  \varrho(r) = \varrho(\sqrt{(X-X_{c})^{2} + (Y-Y_{c})^{2}})
  \noindent
\end{equation}
where the core coordinates $X_{c}$ and $Y_{c}$ are taken as two
additive parameters in the minimization. The adjustments are
generally improved when compared to the original treatments, turning
to lower sizes (in the case of Yakutsk formula) and better
approximation of the density at 600~m. The situation of the most
energetic event of AGASA \cite{ref:Yo} is given in
Table~\ref{tab:AGA1}.

\begin{table}[t]
  \caption{ Localization of the most energetic AGASA event: 
    $\chi^{2}$, $\varrho(600)$, $N_{e}$ and 
    relative core distance from the original localization.
    Age parameter s fitted for A and B functions as 1.05 
    and 0.98, respectively. \label{tab:AGA1}}
\center
  \begin{tabular}{||l||r|r|r|r||}
    \hline
    &
    \multicolumn{1}{c|}{$\chi^{2}$/23} & \multicolumn{1}{c|}{$\varrho(600)$} &
    \multicolumn{1}{c|}{$N_{e}$}  & \multicolumn{1}{c||}{$\Delta r$} \\
    & & \multicolumn{1}{c|}{(1/m$^{2}$)} & \multicolumn{1}{c|}{(10$^{10}$)} &
    \multicolumn{1}{c||}{ (m)} \\
    \hline
    original &  & 892 & 7.84 & 0.00 \\
    \hline
    A -- JNC01 & 7.32 & 598 & 34.81 & 82.44 \\
    B -- JNC02 &  8.24 & 599 & 17.57 & 98.83 \\
    C -- Yakutsk & 10.40 & 561 & 3.01 & 101.15 \\
    D -- Linsley's & 7.51 & 565 & 10.74 & 83.29 \\
    E -- AGASA no. 1 & 6.90 & 580 &  6.43 & 77.10 \\
    F -- AGASA no. 2 & 11.81 & 611 & 8.92 & 60.49 \\
    \hline
  \end{tabular}
\end{table}

\section{Fluorescence, Cerenkov and Radio emissions}

\begin{figure}[t]
  \epsfxsize=5.8cm 
  \epsfbox[0 0 709 765]{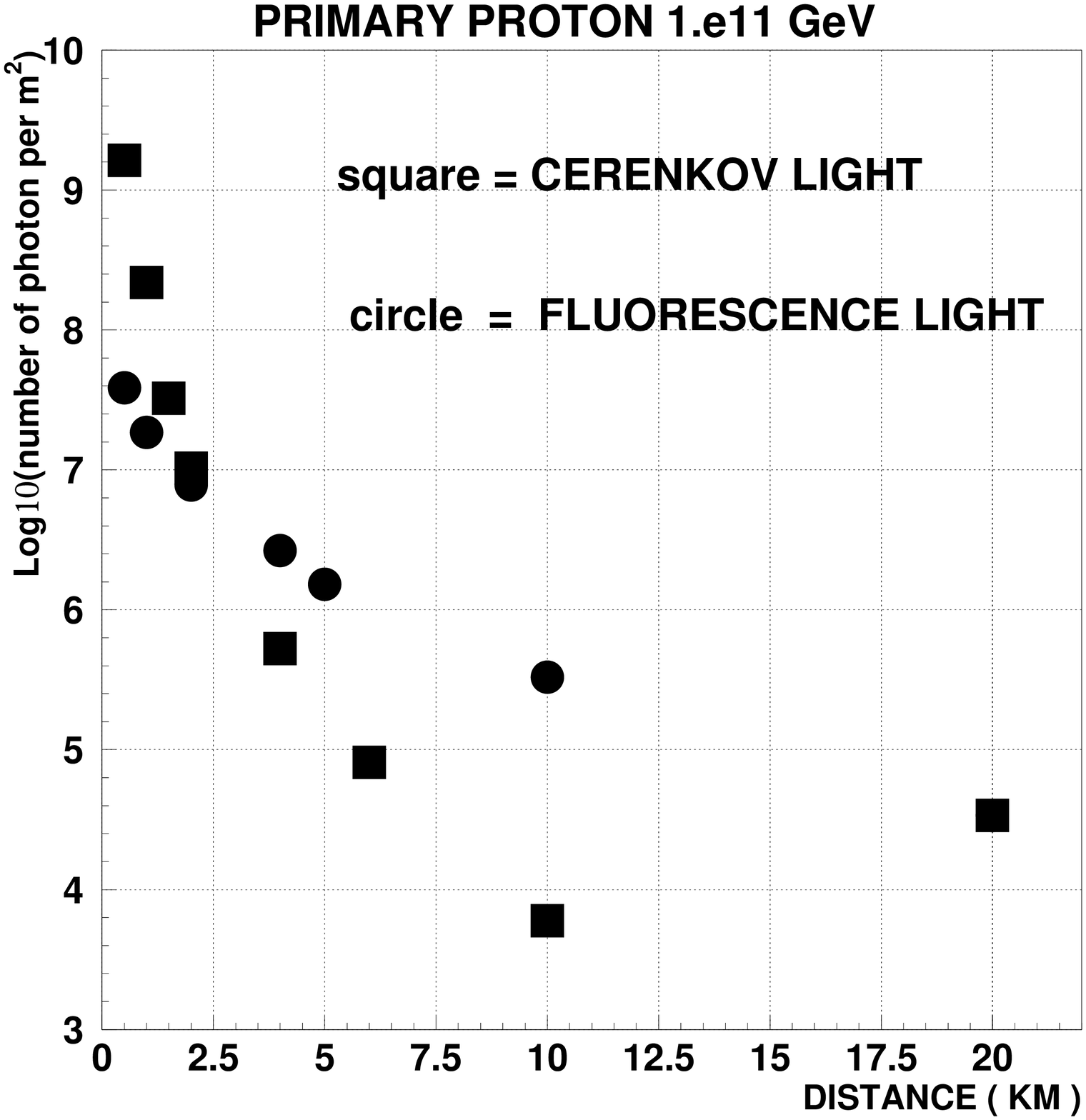} 
  \epsfxsize=5.8cm 
  \epsfbox[0 0 709 765]{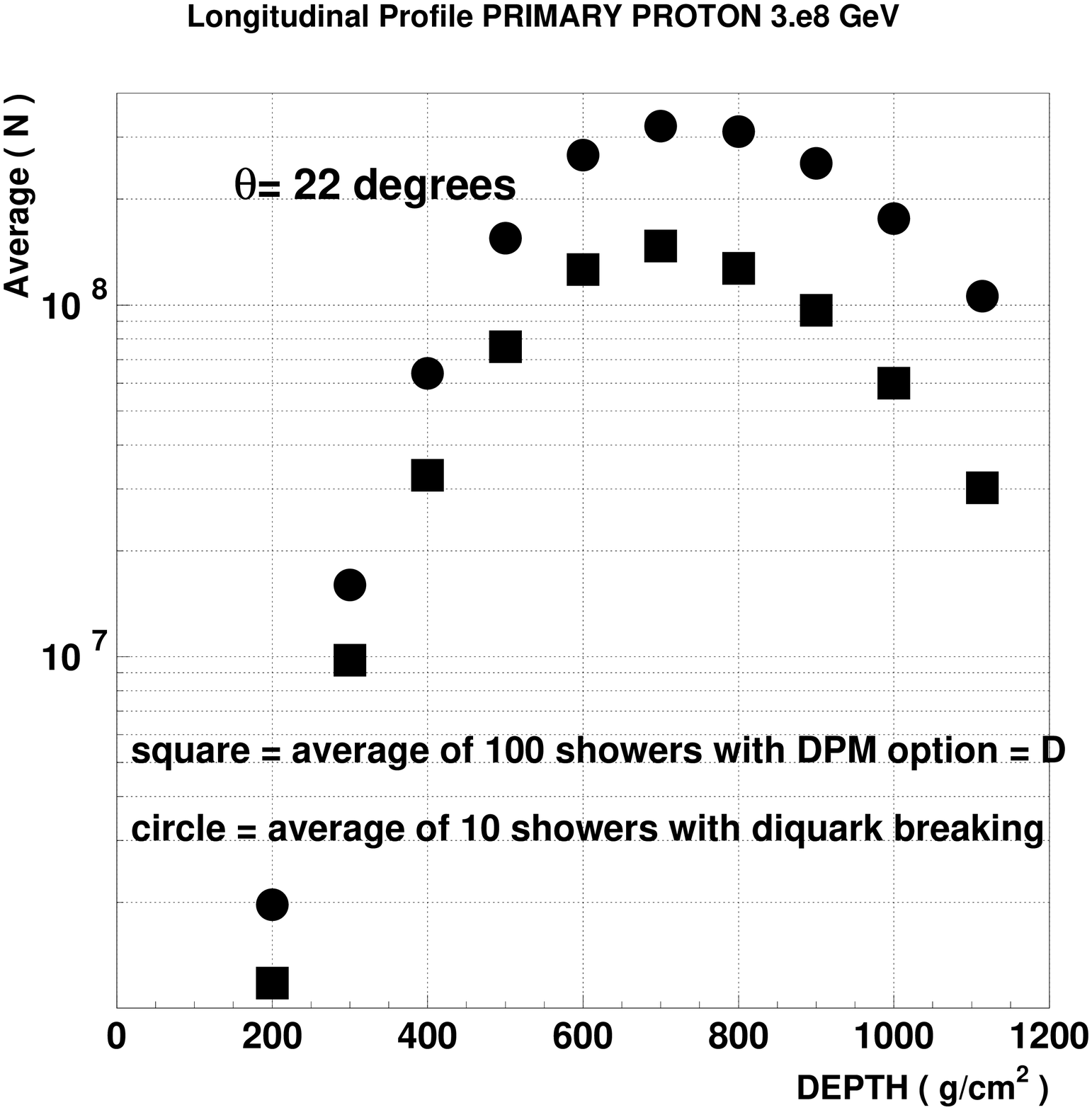} 
    \caption{ 1. Cerenkov and fluorescence light for a proton at $10^{20}$ eV.
      \label{fig:FLUO}  -
      2. Simulation with DPM model with option D (100 showers),
      and diquark breaking (10 showers). \label{fig:DPM} }
\end{figure}

From the longitudinal development, we have developed a method of fast
simulation based on the structural stability of the subshowers (in the
sense of catastroph theory). This is an efficient alternative to the
thinning technique.
The lateral extension of the total amount of light received at a fixed 
distance from the axis is compared for both Cerenkov and fluorescence 
component and shown on Fig.~\ref{fig:FLUO}.\\
The opportunity of CORSIKA which separates positive and negative
particles taught us that a regular negative excess close to $25\%$ is
present in the e.m. component of EAS, allowing the calculation of
radio emission, following Askarian's effect. In Yakutsk experiment,
it is possible to estimate the primary energy from the density 
at $600~m$
 as well as from the Cerenkov component; in AUGER array, the direct
calibration of showers detected simultaneously by the ground array and
the fluorescence detector looks promising.

%
\section{Diquark breaking mechanism and GZK cut off}
The diquark breaking mechanism disturbs strongly the leading particle
effect present in the different models used in cosmic rays. In the
classical form of the dual parton model, the 3 valence quarks of
the proton projectile are separated in a fast diquark and another
valence quark slowed down. The diquark is recombined with one quark of
the sea to produce, the most commonly, an outgoing leader baryon
propagating the energy deeper in the cascade.
The 3 valence quarks separated will 
be recombined in various meson structures in pairs $|u\bar{d}>$,
$|d\bar{u}>$,... or neutral mesons as $1/\sqrt{2}(d\bar{d}-u\bar{u})$.
The configuration with the simultaneous final state for the valence
quarks of $3 \pi^{0}$'s could be especially interesting with a
probability of emergence that we can evaluate from the quark content
and the quark additive model as $1/27$. Such configuration (with
intermediate final states of higher probabilities, one pair of charged
pions and one neutral, one pair of neutral and one charged...) will
transfer a large part of energy to the electromagnetic component and
this energy will be definitely missing for both hadronic and
penetrating components. Remembering that for the same primary energy,
the cascade theory shows that one primary photon produces at maximum,
approximately, two times more electrons, we can expect a large
electron excess for some cascades initiated with diquark breaking.
The longitudinal development calculated for protons of the same energy
of $3\cdot10^{8}$~GeV is compared to a classical development, here the
model HDPM2 with D--option,
on Fig.~\ref{fig:DPM}.  The
electron size at maximum is doubled in the assumption of diquark
breaking and relatively rare recombination simultaneously in $3$
neutral pions \cite{Nik00}.
We have simulated 10$^{4}$ showers, between $10~EeV$ and $50~EeV$, where
the generation, following the primary energy differential spectrum, is
cut by brute force. The showers simulated with axis distributed
randomly are treated following the method of AGASA and the primary
spectrum reconstructed is superimposed on the original one. It appears
an extra component above $50~EeV$, due to the showers including the
diquark mechanism, considered previously, and such artefact could appear
as a GZK violation. With AUGER array, such showers could be recognized
very easily by an energy 3 or 4 times larger from the fluorescence
detector than estimated by the water tank detectors. We observed also
that such artificial component could appear from spacial topological
situations at $50~EeV$; for instance, with an axis falling near one
detector, all the others are hit by densities near the
low density threshold and the answer very unstable of those
surrounding detectors can turn to an important overestimation of the
primary energy.
\section{Conclusions}
The hypergeometric approach gives a better accuracy in the
interpolation of densities at 600--1000~m, a more reliable estimation
of the shower size (when the axis is in the array), a better axis
localization and finally a more correct constraint of the primary
energy. The position of the axis is crucial as the density near
$600~m$ varies as $r^{-4}$ and an uncertainty of $50~m$ on axis turns to
an error of $40\%$ at least on primary energy. Topological
difficulties and also non standard aspect of multiple production needs
probably more attention to be able to rule out definitely the cut
off of GZK.
\section*{Acknowledgments}
 This work has been partly supported (JNC, FC and CLG) by INTAS contract
 1339.


\begin{thebibliography}{99}
\bibitem{ref:JNC99} Capdevielle~J.N. et al.
  in Proc. $26^{th}$ ICRC (Salt Lake City) {1} {1999} {111}.

\bibitem{ref:JNC00} Capdevielle~J.~N., Le~Gall~C., Sanosyan~K.
  in Astroparticle Physics {13} {2000} {259}.

\bibitem{ref:Nag92} Nagano~M. et al.
  in J. Phys. G.: Nucl. Particle Phys. {18} {1992} {423}.

\bibitem{ref:Vis} Vishwanath~P.~R.
  in Proc. $23^{rd}$ ICRC (Calgary), Rapporteur Papers {6} {1993} {384}.

\bibitem{ref:CP} Capdevielle~J.~N., Procureur~J.
  in Proc. 18th ICRC {11} {1983} {307}.

\bibitem{ref:Yo} Yoshida~S. et al.
  in Astroparticle Physics {3} {1995} {105}.

 \bibitem{Saltz00} Saltzberg,~D. et al., 
   preprint hep--ex/0011001, {2000}.

 \bibitem{Aska61} Askaryan,~G.~A., 
   Zh. Exp. Theor. Phys., {41} {616} {1961}.






\bibitem{ref:JNC88} Capdevielle~J.N. et al.
  in J. Phys. G {14} {1988} {503}.

\bibitem{ref:JNC95} Capdevielle~J.N. et al.
  in Proc. $24^{th}$ ICRC (Roma) {1} {1995} {910}.



 \bibitem{Nik00} Nikolsky,~S.~I., Romakhin,~V.~A.,
   Physics of Atomic Nuclei, {63} {10} {1799} {2000}.

\bibitem{ref:AT97} Attallah~R., Capdevielle~J.N., Meynadier ~C. and Szabelski~J.
  in J. Phys. G {22} {1997} {1497};
\end{thebibliography}
\end{document}